# Quantum phase slip junctions

J.E. Mooij* and Yu.V. Nazarov
Kavli Institute of Nanoscience
Delft University of Technology
2628 CJ Delft, The Netherlands
*e-mail: j.e.mooij@tnw.tudelft.nl

*abstract*
For a superconductor, charge and phase are dual quantum variables. In a Josephson junction, particles with charge 2*e* tunnel across an insulating barrier between two superconducting films. Phase-slip is the dual process whereby the phase difference between two superconducting regions changes by $2\pi$ in a short time. Experiments on the resistance of superconducting nanowires at low temperatures were performed to establish whether phase-slip can proceed by quantum tunnelling, but yielded inconclusive results as yet. Büchler et al. investigated finite-length wires theoretically and conclude that quantum phase slip (QPS) can be a coherent process as in a Josephson junction. In this paper we demonstrate that, if it exists, coherent QPS is the exact dual to Josephson tunneling. We use the duality to predict kinetic capacitance and a sharp resonance in narrow wires. Biased resistively and driven at high frequency, QPS junctions should exhibit current plateaus of interest for a fundamental standard.

A phase slip event in a thin superconducting wire occurs on the scale of the coherence length. Phase slip by thermal activation [1] is observed as a resistive tail below the critical temperature. In wires with diameter below 10 nm and very high resistance, the energy barrier is small enough that phase slip by quantum tunnelling can be expected [2,3]. Wires of Mo-Ge deposited on suspended carbon nanotubes, studied in recent years by Bezryadin and co-workers [3,4,5] yield results that compare well with microscopic calculations of phase-slip rates by Zaikin and collaborators [6,7]. All experiments consisted of passing a small DC current through the sample and measuring the voltage. Since each phase-slip event in the presence of a current $I$ releases an energy $I\Phi_0$ where $\Phi_0 = h/2e$ is the flux quantum, such measurements are dissipative. Unambiguous experimental evidence of *coherent* QPS is still absent. In their latest paper [5] Bezryadin and colleagues conclude that their results can be described with thermally activated phase slip for wires with larger cross-section and as mesoscopic diffusive normal metal conductors for the weaker wires. The theoretical analysis is complicated by the fact that the behaviour of the bosonic superfluid in which QPS takes place may be overshadowed by the fermionic effects of localization and interaction. The superconducting energy gap in the wire may be suppressed and quasiparticles may be generated. However, we are not aware of any reason that would forbid QPS to be a physical reality. We thus assume that coherent QPS may take place and that it is characterized by a transition amplitude $E_S/2$. Our predictions derived from this assumption should then be helpful to provide the experimental test of QPS existence. Based on WKB-type estimates [3] and microscopic calculations [7] we assume that the transition amplitude for QPS in practical superconducting nanowires of 1 μm length can be as high as $E_S/h = 150$ GHz. Wires in which significant QPS occurs have large kinetic inductance $L'$ and small capacitance $C'$ per unit length. The plasmon phase velocity $c_p = (L'C')^{1/2}$ is of order $10^5$ m/s [8,9] and the characteristic impedance $Z_c = (L'/C')^{1/2}$ is of order 100 kΩ. Zaikin et al. [6] expect frequent QPS when $Z_c > R_q$, with $R_q = h/4e^2 = 6.45$ kΩ. Büchler et al. [10] made a next step in the understanding of QPS by considering a wire of finite length as a circuit element. If the rest of the circuit is modelled by a parallel resistance $R_p$, they predict that the wire will exhibit a quantum phase transition at $R_p = R_q$, being superconducting at $R_p < R_q$ and insulating otherwise. As predicted by Schmid [11], a Josephon junction displays qualitatively the same behaviour. Büchler et al. draw the conclusion that the low-energy physics of QPS reduces to that of a Josephson junction in the same circuit.

We explicitly demonstrate in this paper that the relation between Josephson and QPS in circuits is more intriguing. They are dual to each other with respect to the exchange of the canonically conjugated quantum variables, phase and charge. This duality is *exact* in contrast to the widely known [12] *approximate* self-duality of Josephson junction circuits. Results of Ref. [8] are reproduced by employing exact duality and approximate self-duality. We use exact duality to describe the dynamical response of QPS in the limit of strong phase slip where we reveal very favourable conditions for the observation of Bloch-type oscillations and of a resonance at the 'plasma' frequency.

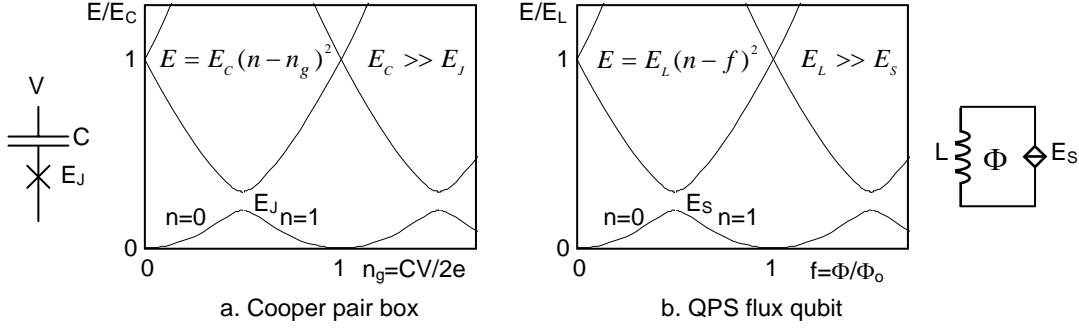

figure 1. Circuit and energy dependence for Cooper pair box and QPS qubit. The diamond-shaped symbol in the QPS qubit circuit represents the quantum phase-slip process.

The duality is explicated in two steps. We first compare two simple circuits (figure 1). One is the Cooper pair box [13] with a Josephson junction that is voltage-biased through a capacitor. In the other we introduce the QPS junction, an element that represents the phase slip process with strength $E_S$, in a closed loop with an inductor. This is the QPS flux qubit proposed in [14], but not yet realized in practice. In the Cooper pair box the *charging* energy depends parabolically on the induced charge $n_g=CV/2e$, for each integer value of the Cooper pair number $n$. The charging energy scale is given by $E_C=(2e)^2/2C$. Josephson coupling $E_J$ mixes states with $n$ and $n+1$, lifting degeneracy at half-integer values of $n_g$. The level splitting at this point equals $E_J$ provided $E_J \ll E_C$. This is described by the following Hamiltonian:

$$H_{JJ} = E_C(n-n_g)^2 - (\frac{E_J}{2}\sum_n |n+1\rangle\langle n| + h.c.) \qquad (1)$$

In the QPS qubit on the right of figure 1 the *inductive* energy depends parabolically on the applied flux with $f=\Phi/\Phi_o$ at each integer $n$ that now represents the fluxoid number in the loop. The scale for the inductive energy is $E_L=\Phi_o^2/2L$. QPS coupling mixes states with adjacent fluxoid numbers and lifts degeneracy at half-integer values of $f$. The level splitting at this point equals $E_S$, provided $E_S \ll E_L$. This implies the following Hamiltonian:

$$H_{QPS} = E_L(n-f)^2 - (\frac{E_S}{2}\sum_n |n+1\rangle\langle n| + h.c.) \qquad (2)$$

The diagram as drawn for the Cooper pair box is only valid when $E_C \gg E_J$, where charge is the relevant quantum number. In analogy, the diagram for the QPS qubit is only valid when $E_L \gg E_S$, the limit of weak phase-slip. Here phase is the relevant quantum number. The above Hamiltonians are obviously equivalent with respect to the exchange

$$E_L \Leftrightarrow E_C; E_J \Leftrightarrow E_S; n_g \Leftrightarrow f \qquad (3)$$

At this stage we have not yet made use of the fact that charge and phase are canonically conjugated quantum variables.

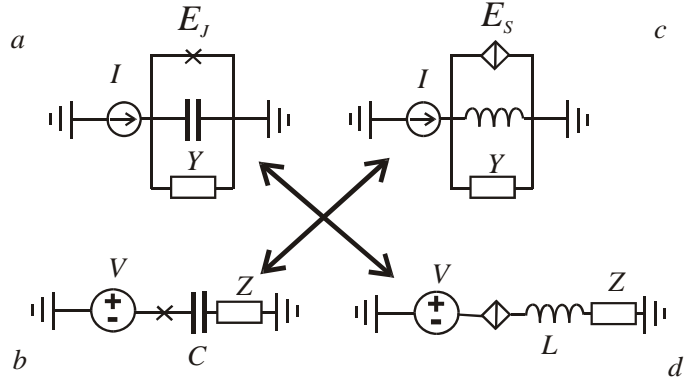

Figure 2. a. Current-biased Josephson junction; b. Voltage biased Josephson junction; c. Current-biased QPS junction; d. Voltage-biased QPS junction. Circuit a. is the exact dual of circuit d., circuit b. is the exact dual of circuit c.

We will now derive the exact duality from this fact. We start by considering a Josephson junction in a linear circuit with either a voltage or a current source (figure 2). Any linear circuit can be presented by an equivalent (frequency-dependent) resistor in series with (for voltage bias) or parallel to (for current bias) the junction. A quantum variable describing the circuit is either phase across the junction or the continuous number of Cooper pairs transferred. Those are canonically conjugated variables so that the corresponding operators satisfy $[\hat{q},\hat{\phi}] = -i$. In the spirit of the Caldeira-Legett approach [15], the circuit can be described with the following Hamiltonian

$$\hat{H} = E_C \hat{q}^2 - E_J \cos\hat{\phi} + \hat{H}_{env} + \hat{H}_{coupling} \qquad (4)$$

Here, $\hat{H}_{env}$ represents boson-like environment modes, and the coupling term $\hat{H}_{coupling}$ is different for voltage or current bias,

$$\hat{H}_{coupling} = \begin{bmatrix} \dfrac{\Phi_0}{2\pi}(I - \hat{I}_r)\hat{\phi} \text{ for current bias} \\ -2e(V - \hat{V}_r)\hat{q} \text{ for voltage bias} \end{bmatrix} \qquad (5)$$

The operators $\hat{I}_r, \hat{V}_r$ present current (voltage) fluctuations in the effective resistor and are linear combinations of the environment bosons. The coefficients in these linear combinations are fixed to reproduce the response function of the environment:

$$\hat{I}_{sh}(\omega) = \dfrac{\hbar}{2e}(-i\omega)Y(\omega)\hat{\phi}(\omega)$$
$$\hat{V}_{sh}(\omega) = 2e(-i\omega)Z(\omega)\hat{q}(\omega) \qquad (6)$$

$Z(\omega)(Y(\omega))$ being the frequency-dependent impedance (admittance) of serial (parallel) equivalent resistor.

Similarly, a QPS junction in a circuit is described by

$$\hat{H} = \dfrac{E_L}{(2\pi)^2}\hat{\phi}^2 - E_S \cos(2\pi\hat{q}) + \hat{H}_{env} + \hat{H}_{coupling} \qquad (7)$$

Let us now consider the effect of the canonical transformation

$(\hat{q}, \hat{\phi}) \to (-\hat{\phi}/2\pi, 2\pi\hat{q})$ on the Hamiltonian (7). Obviously, that transformation does not change the commutation relations. We see that it transforms the QPS Hamiltonian into the Josephson Hamiltonian with the following parameters:

$$E_S \to E_J; E_L \to E_C; I \leftrightarrow R_q^{-1}V; Y(\omega) \leftrightarrow R_q^{-1}Z(\omega) \tag{8}$$

Double-sided arrows mean that the transformation exchanges current and voltage bias and series and parallel resistors (Fig. 2). This is the main exact result of our work. These duality relations allow us to exactly map any known result concerning transport characteristics of Josephson junctions in a circuit to dual transport characteristics of QPS junctions in the dual circuit.

The circuits of figure 2 were in the regimes $E_C \gg E_J$ and $E_L \gg E_S$. Now let us turn to the opposite regimes for the above Hamiltonians. For a Josephson junction, this limit is achieved at $E_J \gg E_C$ where the phase $\phi$ is a well-defined variable. Of particular interest is the dynamics of the classical Josephson junction as described by the resistively shunted junction model

$$I(t) = I_c \sin\phi + \frac{\Phi_0}{2\pi}\left(C\frac{d^2\phi}{dt^2} + \frac{1}{R}\frac{d\phi}{dt}\right) \tag{9}$$

that describes a motion in a slanted sinusoidal ('washboard') potential. Time-independent solutions correspond to the zero-voltage state of the junction where the phase is trapped in one of the potential minima. These minima exist provided the drive current $I$ does not exceed the critical current of the junction $I_c = 2\pi E_J/\Phi_o$. The dual counterpart of this equation for the voltage-biased QPS junction reads

$$V(t) = V_c \sin(2\pi q) + 2e\left(L\frac{d^2q}{dt^2} + R\frac{dq}{dt}\right) \tag{10}$$

This expression is valid in the limit of large phase-slip amplitudes $E_S \gg E_L$ where the charge is a well-defined quantum variable. Here we have a washboard potential in the *charge* variable. The charge may be trapped in potential minima resulting in an insulating zero-current state provided the bias voltage $V$ does not exceed the critical voltage

$$V_c = \frac{2\pi E_S}{2e} \tag{11}$$

For the Josephson junction in the classical regime the response to small signals follows the kinetic inductance $L_{kin} = \Phi_o/(2\pi I_c \cos\phi)$. Similarly for QPS junctions the response is according to a kinetic capacitance

$$C_{kin} = \frac{2e}{2\pi V_c \cos 2\pi q} \tag{12}$$

The geometry of Josephson tunnel junctions implies a shunt capacitance. The combination of kinetic inductance and shunt capacitance leads to a resonance at the so-called plasma frequency $\omega_p = (2E_J E_C)^{1/2}$. Because $E_J$ scales with the area and $E_C$ scales inversely with it, $\omega_p$ is independent of area for a homogeneous barrier. For a QPS junction, the natural geometry is a wire with considerable kinetic inductance. Now, the resonance provided by kinetic capacitance (12) and inductance has a frequency

$$\omega_p = \sqrt{2E_S E_L} \tag{13}$$

As $E_S$ is proportional and $E_L$ is inversely proportional to the wire length the plasma frequency is constant when the length is changed. However, in a lithographically fabricated nanowire it will be easy to lower $\omega_p$ by adding a wider section of wire that lowers $E_L$ without increasing $E_S$ (figure 3). It is interesting to note that moving from long wires to short wires the ratio $E_S/E_L$ decreases in the same way as the decrease of $E_J/E_C$ when going from large to small area Josephson junctions. Whereas in the latter case a transition is made from superconducting to insulating response, the wires may be insulating for long lengths and recover superconductivity at shorter lengths.

The high-frequency environmental impedance of a submicron junction is ususally determined by the geometric capacitance and inductance of nearby wiring and without special precautions has a value around 300 $\Omega \ll R_q$. For a QPS junction this natural impedance leads to a high quality factor of the plasma oscillations given by

$$Q_{QPS}^2 = \beta_L = 2\pi \frac{V_c}{2e} \frac{L}{R^2} = 2\pi^2 \frac{E_S}{E_L} \left(\frac{R_q}{R}\right)^2 \qquad (14)$$

Here $\beta_L$ is the equivalent of the well-known McCumber parameter $\beta_C$ for Josephson junctions. Employing duality, we predict a strongly non-linear hysteretic response to relatively small resonant signals $V_{AC}/V_C \sim 1/Q$. The dual effect was recently investigated for a precision quantum measurement with Josephson junctions [16].

The most far-reaching application of QPS junctions might be in a fundamental standard of current, dual to the well-known Josephson voltage standard. Equation (10) can be used to calculate the transport in the series circuit of QPS junction, inductance and resistance. When an RF voltage is applied, the equivalent of Shapiro steps will occur in the form of plateaus at constant current levels

$$I_n = n2e\nu \qquad (15)$$

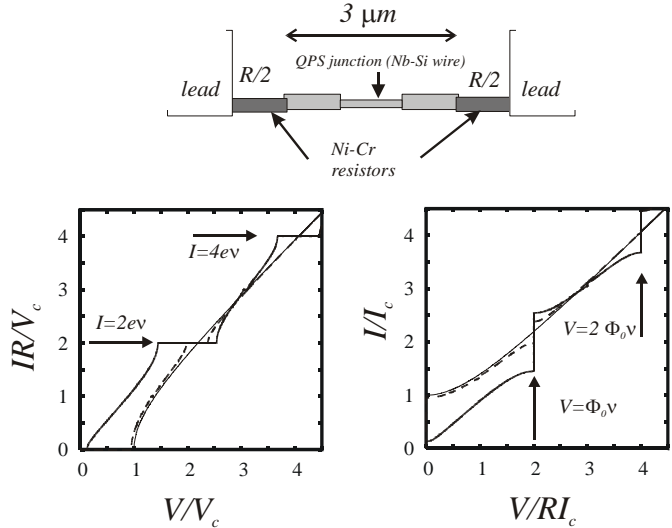

Figure 3. Shapiro steps at constant current in a QPS junction. Above: The junction must be embedded in a highly resistive environment. Parasitic capacitances are sufficiently small for the design given. Left: QPS junction. The parameters correspond to $Q=0.5$, $\omega_{AC} = \omega_p$, $V_{ac}/V_c =$ 0,1,5 for the thick solid, dashed and thin solid curves respectivily. With $E_S/h = 120$ GHz and $E_L/h = 30$ GHz the plasma frequency is $\omega_p/2\pi = 85$ GHz and the critical voltage is $V_c = 1.56$ mV; $R = 115$ k$\Omega$. Right: Voltage Shapiro steps in the dual Josephson junction circuit.

where ν is the RF frequency. For $n=1$ and $\nu = 80$ GHz, as used in Josephson voltage standards, the current level would be 26 nA. High $Q_{QPS}$ is unfavourable for this application since it gives rise to a hysteretic current-voltage characteristic and unrealistically high frequencies $\sim Q_{QPS}$ and modulation amplitudes $\sim Q_{QPS}^2$. To achieve a practical device, the circuit should be made non-hysteretic by the use of a sufficiently high series resistance. As $E_S/E_L$ cannot be smaller than about 4 to guarantee that charge, as the relevant quantum variable, is well-defined, the series resistance has to be at least 60 kΩ. In figure 3 we give current-voltage characteristics for a circuit with $Q=0.5$, driven at the plasma frequency. Realistic parameters are given in the figure. To fabricate the QPS wire as well as two series resistors of 60 kΩ each, without shunting by parasitic capacitances, is challenging but seems possible with state of the art electron beam lithography. Many devices could be put in parallel on a chip.

The QPS junction at $E_S \gg E_L$ and the Josephson junction at $E_C \gg E_J$ have in common that the charge is the relevant quantum variable. For Josephson junctions in the charging regime, Bloch oscillations were predicted [17] and observed [18]. This system is often considered as dual to the classical Josephson junction, because of the cross-over from phase to charge as quantum variable. However, there is no strict duality as shown for Josephson and QPS. Interestingly, the weak QPS regime with $E_L \gg E_S$ and the classical Josephson junction share the phase as the quantum variable.

In conclusion, we have shown that quantum phase slip is the exact dual to Josephson tunnelling. This duality leads to clear predictions that can be tested experimentally. The existence of coherent quantum phase slip can unequivocally be established by such experiments. Quantum phase slip junctions may have interesting device applications as resonators and as fundamental current standards.

*acknowledgements*
We thank L. Levitov, L. Glazman, T.P. Orlando, and C.J.P.M.Harmans for discussions. Our research is supported by FOM, NanoNed and EuroSQIP.

references

[1] Tinkham, M. *Introduction to Superconductivity* 2$^{nd}$ edn, p.288, (New York, McGraw-Hill, 1996).
[2] Giordano, N. Evidence for macroscopic quantum tunneling in one-dimensional superconductors. Phys.Rev.Lett. 61, 2137-2140 (1988).
[3] Lau, C.N., Markovic, N., Bockrath, M., Bezryadin, A. & Tinkham, M. Quantum phase slips in superconducting nanowires. Phys. Rev. Lett. 87, 217003 (2001).
[4] Bezryadin, A., Lau, C.N. & Tinkham, M. Quantum suppression of superconductivity in ultrathin nanowires. Nature 404, 971-974 (1999).
[5] Bollinger, A.T., Rogachev, A. & Bezryadin, A. Coulomb blockade in the insulating regime of short superconducting nanowires. arXiv:cond-mat/0508300
[6] Zaikin, A.D., Golubev, D.S., van Otterlo, A. & Zimanyi, G.T. Quantum phase slips and transport in ultrathin superconducting wires. Phys. Rev. Lett. 78, 1552-1555 (1997).
[7] Golubev, D.S. & Zaikin, A.D. Quantum tunneling of the order parameter in superconducting nanowires. Phys. Rev. B 64, 014504 (2001).
[8] Büchler, H.P., Geshkenbein, V.B. & Blatter, G. Quantum fluctuations in thin superconducting wires of finite length. Phys. Rev. Lett. 92, 067007 (2004).
[9] Kulik, I.O. Frequency dependence of the penetration depth of an electric field in a superconductor. Fiz. Nizk. Temp. 5, 1391 (1979), English translation: Sov. J. Low Temp. Phys. 5, 656-662 (1979)
[10] Mooij, J.E. & Schön, G. Propagating plasma mode in thin superconducting filaments. Phys. Rev. Lett. 55, 114-117 (1985).
[11] Schmid, A. Diffusion and localization in a dissipative quantum system. Phys. Rev. Lett. 51, 1506-1509 (1983).
[12] Schön, G. & Zaikin, A.D. Quantum coherent effects, phase transitions, and the dissipative dynamics of ultra small tunnel junctions. Phys. Rep. 198, 237-412 (1990).
[13] Bouchiat, V., Vion, D., Joyez, P., Esteve, D. & Devoret, M.H. Quantum coherence with a single Cooper pair. Phys. Scripta T 76, 165-170 (1998).
[14] Mooij, J.E. & Harmans, C.J.P.M. Phase-slip flux qubits. New J. of Phys. 7, 219 (2005).
[15] Caldeira, A.O. & Leggett, A.J. Quantum tunnelling in a dissipative system. Ann. Phys. 149, 374-456 (1983).
[16] Siddiqi, I. et al., Direct observation of dynamical bifurcation between two driven oscillation states of a Josephson junction. Phys. Rev. Lett. 94, 027005 (2005).
[17] Likharev, K.K. & Zorin, A.B. Theory of the Bloch-wave oscillations in small Josephson junctions. J. Low Temp. Phys. 59, 347-382 (1985).
[18] Kuzmin, L. & Haviland, D.B. Observation of the Bloch oscillations in an ultrasmall Josephson junction. Phys. Rev. Lett. 67, 2890-2893 (1991).